# Coalition Structure and Polarization in Parliamentary Voting Networks

## Evidence from the Italian Parliament

Francesca Collu[1,2]
Antonio Scala[2,3]
Emilia La Nave[2]

[1] Statistics Department, Sapienza University, Roma, RM, Italy
[2] Applico Lab, CNR-ISC, Roma, RM, Italy
[3] Centro Enrico Fermi, Roma, RM, Italy

PACS: 89.65.-s; 89.75.Fb

MSC: 05C; 62H30; 62P25

JEL: D72; D85; C63

**Corresponding author:**
Antonio Scala
Email: antonio.scala@cnr.it

# Highlights

- Voting similarity networks reveal latent coalition structure beyond formal labels
- A reproducible network framework links individual actions to system-level patterns
- Network polarization depends on coalition structure rather than coalition size
- Group coherence and individual alignment decouple during political transitions
- System-wide polarization arises from interaction structure, not dominant groups

# Coalition Structure and Polarization in Parliamentary Voting Networks

Evidence from the Italian Parliament

true

true

true

# Abstract

Ensuring legislative accountability in multi-party systems requires quantitative tools that reveal actual voting behavior beyond formal party affiliations. We present a network-based framework for analyzing parliamentary dynamics at multiple scales, capturing coalition structure, group coherence, and individual influence. Applied to over 4 million vote expressions from the Italian Parliament across three government formations (2018-2021), the methodology combines network modularity, voting distance metrics, and betweenness centrality to map the structure of collective decision-making.

Using this framework, we show that system-level polarization, as captured by network modularity, varies systematically with coalition structure rather than coalition size. Technical governments display paradoxically lower global polarization despite broader formal support, reflecting structurally mixed voting patterns rather than unified blocs. On polarizing issues such as immigration, network polarization depends strongly on the fragmentation or cohesion of the opposition, even when the governing coalition votes cohesively. Betweenness analysis reveals that mediator roles are highly concentrated, with only about 2% of parliamentarians acting as structural bridges between communities. Community detection further uncovers implicit coalitions that are not apparent from declared alliances.

The framework relies exclusively on public roll-call data, enabling reproducible analysis and direct applicability to any legislature with transparent voting records. By linking individual voting behavior to emergent system-level structure, the methodology provides quantitative infrastructure for comparative analysis of legislative voting networks and coalition monitoring, enabling systematic assessment of legislative behaviour.

**Keywords:** voting networks, network modularity, polarization, community structure, coalition dynamics, roll-call voting

---

# Author Summary

Understanding who votes with whom in parliaments is crucial for democratic accountability, but traditional analysis often misses the gap between declared political alliances and actual voting behavior. We developed a network-based method that analyzes millions of individual votes to reveal three levels of political power: how parties cluster into voting blocs, how disciplined parties are internally, and which individual politicians act as bridges between groups.

Testing our approach on the Italian Parliament during a turbulent period with three different governments, we discovered unexpected patterns: governments with broad support showed more internal division than partisan coalitions, and a supposedly unifying issue like immigration actually revealed deeper fragmentation. Our method works with any parliament that publishes voting records, providing a tool for citizens, journalists, and researchers to monitor whether representatives vote as they claim.

# 1 Introduction

Ensuring the accountability of legislatures is pivotal for the effective functioning of modern democracies, as they serve as core representative institutions (Carey, 2008). In multi-party parliamentary systems, understanding coalition dynamics requires moving beyond formal party affiliations to analyze actual voting behavior. While parties may declare alliances or opposition roles, their members' votes often reveal more complex patterns of cooperation and conflict. Quantitative tools for analyzing roll-call votes have been developed for almost a century to comprehend the voting engagement of political parties (Rice, 1925), yet traditional approaches often miss the discrepancies between declared political positions and legislative behavior.

From a network-science perspective, the challenge is not the lack of voting data, but the absence of a general framework that integrates standard network measures into a coherent, multi-scale description of collective political behavior.

Today, open data provides convenient access to the voting behavior of Members of Parliament (MPs), enabling scrutiny of media commentary and political statements against the actual conduct of individual representatives and declared political coalitions. Network analysis emerges as a crucial instrument for analyzing such phenomena, recognizing that "politics is a matter of relations" (Victor et al., 2017). By representing voting patterns as networks – where nodes are parliamentarians and edges reflect co-voting behavior – we can apply established metrics from network science to reveal coalition structure, party discipline, and individual influence simultaneously.

Network theories have found application in various parliamentary case studies, ranging from analyzing the internal cohesion and inter-party cooperation of Italian MPs (Dal Maso et al., 2014) and Czech MPs (Brabec et al., 2019) to examining the rise of polarization in the U.S. Congress (Andris et al., 2015; Moody and Mucha, 2013). Network approaches have been instrumental in uncovering political and organizational correlations between committees in the U.S. House of Representatives (Porter et al., 2005) and revealing cohesion and coalition formation in the European Parliament (Cherepnalkoski et al., 2016). Moreover, network analysis has proven valuable in understanding the relationships between politics and lobbying (Kim and Kunisky, 2021) and even between parliament and corruption (Colliri and Zhao, 2019; Ribeiro et al., 2018).

While existing studies have successfully applied individual network metrics to specific legislatures or periods, network analysis is most often used descriptively, rather than as a transferable methodological framework operating simultaneously across multiple analytical scales.

As a consequence, what remains underdeveloped is a comprehensive framework that: (1) operates at multiple scales simultaneously – from individual MPs to party systems; (2) captures coalition dynamics across varying government structures (partisan coalitions, technical governments, fragmented systems); (3) uses exclusively public data to ensure reproducibility; and (4) supports continuous monitoring as new votes accumulate. Such a framework would enable comparative studies across parliaments and serve as quantitative infrastructure for democratic accountability tools.

# Our Contribution

In this paper, we present a network-based framework that addresses these gaps by combining three complementary approaches: network modularity to reveal coalition structure, voting distance metrics to quantify party coherence, and betweenness centrality to identify influential mediators. We validate this methodology using the Italian Parliament during the 18th legislature (2018-2021), a period characterized by exceptional political turbulence with three distinct government formations. This context provides an ideal natural experiment: the same parliament, the same parties, but radically different coalition structures – from a partisan right-populist coalition (Conte I) to a left-populist alliance (Conte II) to a technocratic "grand coalition" (Draghi).

Leveraging open-access databases from the Italian Parliament, we analyze over 4 million individual vote expressions across both chambers (Camera and Senato) during the first 50 days following each cabinet formation, when coalition dynamics are most fluid. We demonstrate how the framework reveals patterns invisible to traditional analysis:

- **Coalition structure paradox**: Technical governments with broad formal support (Draghi) show *lower* network modularity than partisan coalitions, suggesting parties maintain distinct identities through selective defection rather than merging into a unified bloc.

- **Opposition matters for polarization**: On polarizing topics like immigration, network separation depends more on opposition structure than government cohesion – a fragmented opposition in Conte I produced lower modularity than the supposedly more divided Conte II period.

- **Concentrated mediator power**: Only 2.1% of MPs exhibit significant betweenness centrality, revealing that bridge-building power is highly concentrated among specific individuals, often party-switchers and veteran politicians.

- **Implicit coalitions**: Community detection exposes undeclared alliances, such as 73% of one governing party's senators (LSP-PSDAZ) consistently voting with the opposition during Draghi, despite formal government membership.

Beyond these empirical findings, our primary contribution is methodological: we demonstrate that this framework can be applied to any parliament with public roll-call data. The exclusive reliance on voting records – without requiring text analysis, media data, or proprietary information – ensures reproducibility and enables real-time monitoring as new votes become available. All metrics are formally defined, and our data collection strategy using SPARQL endpoints can be adapted to other parliamentary systems adopting open data standards.

# Structure of the Paper

The paper proceeds as follows. Section 2 describes the data collection process and defines the network construction, voting distance metrics, and community detection algorithms. Section 3 presents results organized around the framework's multiple scales:

overall network properties, coalition structure evolution, party coherence dynamics, and individual mediator roles, with a focused case study on immigration votes. Section 4 discusses theoretical implications for understanding multi-party systems, addresses limitations, and outlines the framework's broader applicability. Section 5 concludes by reflecting on how this methodology contributes to legislative transparency and democratic accountability.

# 2 Materials and Methods

We provide an overview of the analyzed database and the metrics used to extract information on the voting behavior of Political Groups (PGs). Section 2.1 details the organization of the database, Section 2.2 describes the construction of co-voting networks, Section 2.3 outlines the metrics for defining cohesion and voting distances among MPs and PGs, and Section 2.4 discusses the framework's generalizability.

## 2.1 Voting Dataset

The dataset utilized in this study encompasses comprehensive information on votes in the Italian Parliament during the 18th legislature. Specifically, we analyze votes cast between March 23rd, 2018, and September 23rd, 2021, spanning a three-and-a-half-year period across three cabinet formations (Conte I, Conte II, and Draghi).

These data are openly accessible and can be obtained from the parliament's open data SPARQL endpoints: http://dati.camera.it/sparql for the Camera (lower house) and https://dati.senato.it/sito/23 for the Senato (upper house). We executed SPARQL queries to retrieve datasets containing personal details of Parliament members (MPs), bills, Political Groups (PGs), and party switchings. Through targeted queries and automated collection strategies, we successfully compiled over 4 million individual vote records. Technical details on data collection procedures, including workarounds for server limitations and database schema, are provided in Appendix A.

The collected data were organized into a relational database using the Python library SQLAlchemy (Bayer, 2012) and MySQL. Each vote record includes: MP identifier, vote expression (yes/no/abstain/absent), bill identifier, date, and parliamentary group affiliation at the time of voting. This structure enables tracking of individual voting behavior, party switching events, and temporal evolution of coalitions.

For the main analysis, we focus on the first 50 days following each cabinet formation, as this period captures coalition stabilization dynamics while maintaining comparable sample sizes across governments. This window encompasses approximately 160 roll-call votes per chamber per cabinet, totaling over 400,000 individual vote expressions in the focal analysis.

## 2.2 Voting Networks

We represent parliamentary voting behavior through weighted networks where nodes are MPs and edges reflect co-voting patterns. For analytical purposes, we construct networks for members of the Italian Parliament who participated in the same roll-call votes. Each network possesses the following characteristics (Newman, 2003; Wasserman and Faust, 1994):

- $N$ nodes representing deputies or senators participating in the same roll-call votes.

- $|E|$ edges (with $|E| \in [0, N(N-1)/2]$) among MPs who have displayed the same voting behavior at least once (i.e., voted in favor, against, or abstained).
- The weight $w_{ij} \in [0, 1]$ indicating the frequency with which two members $i$ and $j$ (where $i \neq j$) voted the same way. Specifically, $w_{ij} = 1$ if $i$ and $j$ voted identically on all considered roll calls, and $w_{ij} = 0$ if they never voted the same way. For the diagonal case $i = j$, we set $w_{ii} = 0$.

The matrix $W = \|w_{ij}\|$ defines the co-voting network. The weights $w_{ij}$ form the weighted adjacency matrix, describing a simple, undirected graph due to its symmetry with zero diagonal. The adjacency matrix $a_{ij}$ (defined as $a_{ij} = w_{ij}$ if $w_{ij} > 0$, and $a_{ij} = 0$ otherwise) indicates whether two nodes are connected by an edge.

The overall cohesion of the graph is described by the network density:

$$d = \frac{\sum_{i<j} w_{ij}}{N(N-1)/2}, \quad d \in [0, 1]$$

Graph theory provides several measures to identify node importance or prominence (Wasserman and Faust, 1994). The **degree centrality** for a weighted graph is given by:

$$C(N_i) = \frac{\sum_j w_{ij}}{N-1}$$

This measures the node strength $s_i = \sum_j w_{ij}$ normalized by the maximum number of possible connected edges.

Another essential property is a node's position in connecting other nodes, known as the **betweenness index**:

$$C_B(N_i) = \frac{\sum_{j<k} \frac{g_{jk}(N_i)}{g_{jk}}}{\frac{(N-1)(N-2)}{2}}$$

Here, $g_{jk}$ is the number of shortest paths linking nodes $j$ and $k$, and $g_{jk}(N_i)$ is the number of shortest paths constrained to contain node $N_i$. High betweenness indicates MPs who serve as bridges between different voting communities.

## Community Detection

One of our main objectives is to understand and interpret the political behavior of MPs through community structure – a property commonly observed in real networks where nodes divide into groups with dense internal connections and sparser connections between groups (Newman and Girvan, 2004). We employ modularity optimization, a well-established method for detecting communities (Newman, 2004; Newman and Girvan, 2004). Throughout this work, network modularity is interpreted as a system-wide measure of voting polarisation emerging from all pairwise voting relations, rather than as a direct proxy for government – opposition separation.

The modularity $Q$ for weighted graphs is given by:

$$Q = \frac{1}{2W} \sum_{i,j} \left[ w_{ij} - \frac{k_i k_j}{2W} \right] \delta(C_i, C_j)$$

where $W$ is the total weight, $w_{ij}$ is the fraction of similar votes between nodes $i$ and $j$, $k_i$ and $k_j$ are the weighted degree centralities of nodes $i$ and $j$, and $\delta(C_i, C_j)$ equals 1 if nodes $i$ and $j$ belong to the same community and 0 otherwise.

For detecting the communities, we employ the Greedy modularity optimization algorithm implemented by the Python library NetworkX (Hagberg et al., 2008). This algorithm starts with each node in its own community and iteratively joins pairs of communities that most increase modularity until no such pair exists. To assess robustness, we repeated the modularity optimisation multiple times with different initial conditions and random seeds, obtaining identical community assignments in all cases: this indicates that the detected partitions correspond to stable optima rather than algorithmic artefacts.

## 2.3 Voting Distances and Coherence Metrics

To assess policy locations and group cohesion, we define pairwise distances among voters following established approaches in legislative studies (Enelow and Hinich, 1984; Mokken and Stokman, 1985). Consider $N$ voters participating in $M$ roll calls on a set $V = y, n, a$ with three alternatives ("yes," "no," "abstain/absent"). Let $v_p^i \in V$ denote the vote of the p-th voter on the i-th roll call; thus, the vector $v_p \in V^M$ indicates the votes of the p-th voter in $M$ roll calls, while the vector $v^i \in V^N$ represents the votes of $N$ voters in the $i$-th roll call.

Unlike spatial models (Enelow and Hinich, 1984) which explicitly define a policy space along whose dimensions political preferences can be measured, the voting distance approach sacrifices this specific information while retaining details about shared policies. We define the distance between two votes using the following matrix:

| $d(\cdot,\cdot)$ | y | a | n |
|---|---|---|---|
| y | 0 | 1 | 2 |
| a | 1 | 0 | 1 |
| n | 2 | 1 | 0 |

Here, $d(y, y) = d(a, a) = d(n, n) = 0$, $d(y, a) = d(a, y) = d(n, a) = d(a, n) = 1$, and $d(y, n) = d(n, y) = 2$. This can be achieved by identifying $y \rightarrow 1$, $a \rightarrow 0$, $n \rightarrow -1$ and defining $d(v_p^i, v_q^j) = |v_p^i - v_q^j|$.

We thus define the voting distance between two voters over $M$ roll calls as:

$$d(v_p, v_q) = \sum_{i=1}^{M} d(v_p^i, v_q^i) = \sum_{i=1}^{M} |v_p^i - v_q^i|$$

and the distance matrix $D$ among voters having elements $D_{pq} = d(v_p, v_q)$.

Since the maximum distance between two voters over $M$ roll calls is $d_{max} = 2M$, we define a normalized distance $\tilde{d}(v_p, v_q) = d(v_p, v_q)/d_{max}$ between 0 and 1, with corresponding normalized distance matrix $\tilde{D}$.

Suppose voters are partitioned into non-overlapping groups $G = A, B, C, \ldots$ of sizes $N_A, N_B, N_C, \ldots$ with $\sum_{X \in G} N_X = N$. If we order voter indices such that $1, \ldots, N_A$ correspond to voters in group A, then the indices $N_A + 1, \ldots, N_A + N_B$ correspond to voters in group $B$, and so on, the distance matrix $D$ has a block form:

$$D = \begin{pmatrix} D_{AA} & D_{AB} & D_{AC} & \cdots \\ D_{BA} & D_{BB} & D_{BC} & \cdots \\ D_{CA} & D_{CB} & D_{CC} & \cdots \\ \vdots & \vdots & \vdots & \ddots \end{pmatrix}$$

with blocks like $D_{AA}$ describing intra-group distances and $D_{AB}$ describing inter-group distances. The same applies to the normalized distance matrix $\widetilde{D}$.

If $\widetilde{D}_{GG} = 0$ for a group $G$, we regard $G$ as demonstrating perfect group coherency. Denoting the average normalized distance in group $G$ as $a(G, G)$:

$$a(G, G) = \frac{\sum_{p,q \in G : p < q} \widetilde{d}(v_p, v_q)}{N_G(N_G - 1)/2}$$

(notice that we exclude diagonal terms $p = q$), we define the **intra-group coherency measure**:

$$c(G, G) = 1 - a(G, G)$$

such that $c = 1$ when the group is perfectly coherent (i.e., all its MPs vote identically). Similarly, denoting the average normalized distance between two groups as:

$$a(A, B) = \frac{\sum_{p \in A, q \in B} \widetilde{d}(v_p, v_q)}{N_A N_B}$$

we define the **inter-group coherence**:

$$c(A, B) = 1 - a(A, B)$$

To identify incoherent members within a group, we define the average distance $a_p(G)$ of member $p \in G$ to other members of the group as:

$$a_p(G) = \frac{\sum_{q \neq p, q \in G} d(v_p, v_q)}{N_G - 1}$$

and define the **coefficient of location in a group** as the coherence of the member with respect to the group:

$$c_p(G) = 1 - a_p(G)$$

A “loyal” group member p would have $c_p \sim 1$: thus, group members can be ordered by their “loyalty”; the member (or members) with the highest $c_p$ would be the one most coherent with the group’s voting decisions.

# 2.4 Generalizability and Reproducibility

The framework presented here is designed for broad applicability across parliamentary systems. The methodology requires only three data elements available in most parliamentary open data systems: (1) individual vote expressions (yes/no/abstain) with timestamps, (2) MP identifiers enabling tracking across votes, and (3) party or group affiliations at the time of each vote. No additional data – such as bill content, committee assignments, or media coverage – is necessary for the core analysis.

All metrics defined above (co-voting networks, voting distances, coherence measures, modularity) are parameter-free or use standard thresholds from network science literature. The choice of a 50-day window after cabinet formation captures coalition stabilization dynamics while maintaining comparable sample sizes, but any time window can be selected based on specific research questions. The metrics scale naturally from individual MPs to entire party systems without modification.

Our data collection strategy using SPARQL endpoints is directly applicable to parliaments adopting semantic web standards, including legislatures in the UK, France, and the European Parliament. For parliaments with different data formats, the key requirement is programmatic access to vote-level data – whether through APIs, bulk downloads, or web scraping. The database schema (detailed in Appendix A) provides a template for organizing heterogeneous data sources into the relational structure required by our analysis pipeline.

All code and detailed documentation will be made publicly available, enabling full replication and adaptation to other legislative contexts. The exclusive reliance on public roll-call data ensures that the framework can be applied wherever transparent voting records exist, from established democracies to emerging parliamentary systems.

# 3 Results

We organize results around the framework's multiple analytical scales: overall network properties that characterize parliamentary structure (Section 3.1), community detection revealing coalition dynamics (Section 3.2), coherence metrics capturing party discipline evolution (Section 3.3), betweenness centrality identifying influential mediators (Section 3.4), and a focused case study on immigration votes validating the framework on polarizing issues (Section 3.5). Throughout, we compare patterns across the three cabinets (Conte I, Conte II, Draghi) and both chambers (Camera and Senato) to demonstrate how the methodology captures evolving political dynamics.

For clarity of presentation, we associate each Political Group (PG) with consistent acronyms and color codes used throughout all figures. Camera groups: Movimento 5 Stelle (M5S, yellow), Lega Salvini Premier (LSP, green), Forza Italia (FI, blue), Partito Democratico (PD, red), Italia Viva (IV, pink), Fratelli d'Italia (FDI, black), Liberi e Uguali (LEU, orange), and Misto (MISTO, gray). Senato groups follow similar conventions with minor differences (e.g., LSP-PSDAZ includes Partito Sardo d'Azione, IV-PSI includes Partito Socialista Italiano). Complete group compositions and seat distributions are provided in Tables S1-S2 (Supplementary Materials). The acronyms and the colour codes for the parliamentary groups used in the figures are defined in Tables 0a – 0b.

**Table 0a - MPs' Acronyms and color codes (Camera)**

| Group | Acronimo | Colore |
|---|---|---|
| Movimento 5 Stelle | M5S | Yellow |
| Lega Salvini Premier | LSP | Green |
| Forza Italia - Berlusconi Presidente | FI | Blue |
| Partito Democratico | PD | Red |
| Italia Viva | IV | Purple |
| Fratelli d'Italia | FDI | Black |
| Liberi e Uguali | LEU | Dark Red |
| Misto | MISTO | Grey |

**Table 0b - MPs' Acronyms and color codes (Senate)**

| Group | Acronimo | Colore |
|---|---|---|
| Movimento 5 Stelle | M5S | Yellow |
| Forza Italia - Berlusconi Presidente | FI | Blue |
| Partito | | |

Democratico | PD | Red | | Fratelli d'Italia | FDI | Black | | Lega Salvini Premier - Partito Sardo d'Azione | LSP-PSDAZ | Verde | | Per le Autonomie | AUT | Pink | | Italia Viva - Partito Socialista Italiano | IV-PSI | Purple | | Misto | MISTO | Grey |

## 3.1 Network Structural Properties Reveal Coalition Types

We begin by examining fundamental network metrics across the three cabinets to establish how graph structure reflects political organization. Figure 1 displays the co-voting networks during the first month after each cabinet formation, representing approximately 160 roll-call votes in both Camera and Senato. Visual inspection immediately reveals distinct patterns: Conte I and Conte II show clear bipartite structure with two dense communities connected by sparse edges, while Draghi exhibits a more homogeneous core with peripheral isolated nodes.

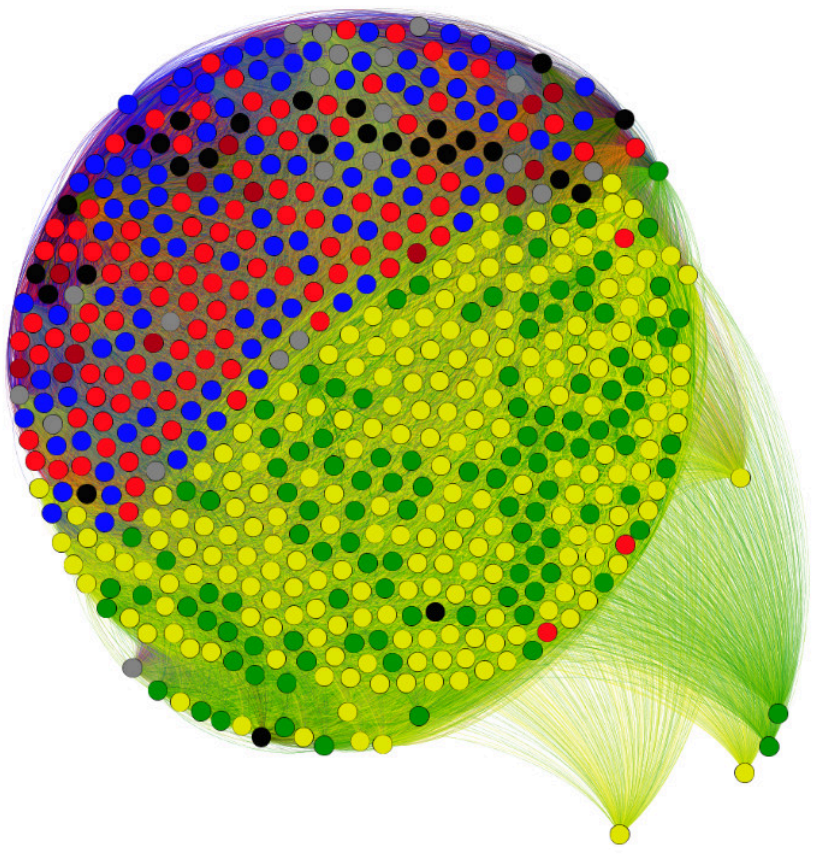

**(a)** Camera, *Conte I* cabinet.

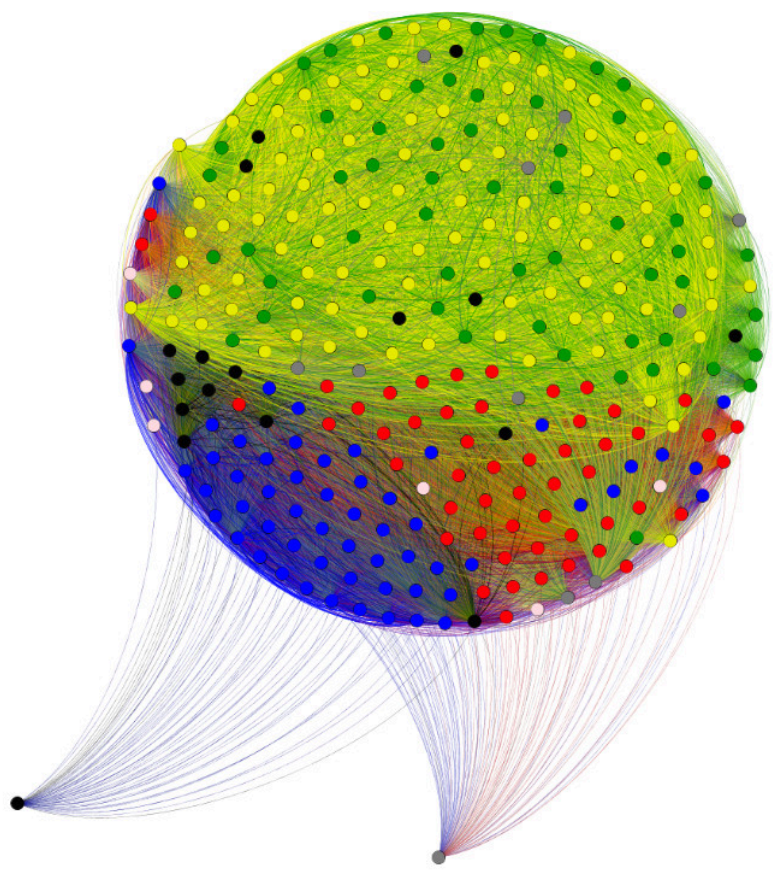

**(b)** Senato, *Conte I* cabinet.

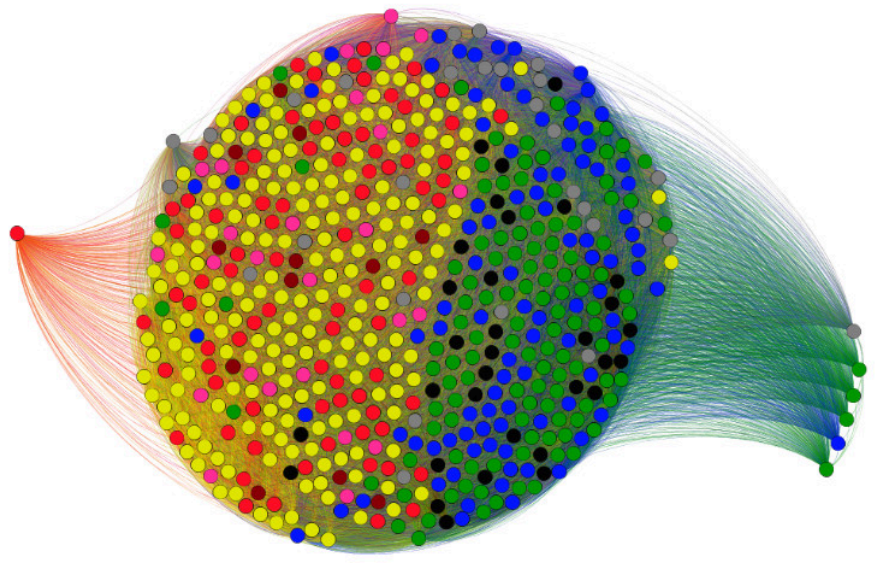

**(c)** Camera, *Conte II* cabinet.

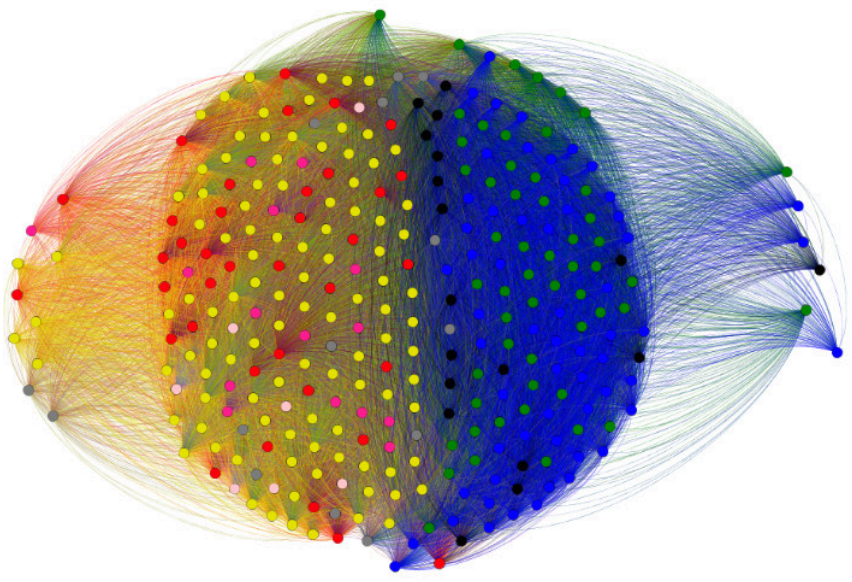

**(d)** Senato, *Conte II* cabinet.

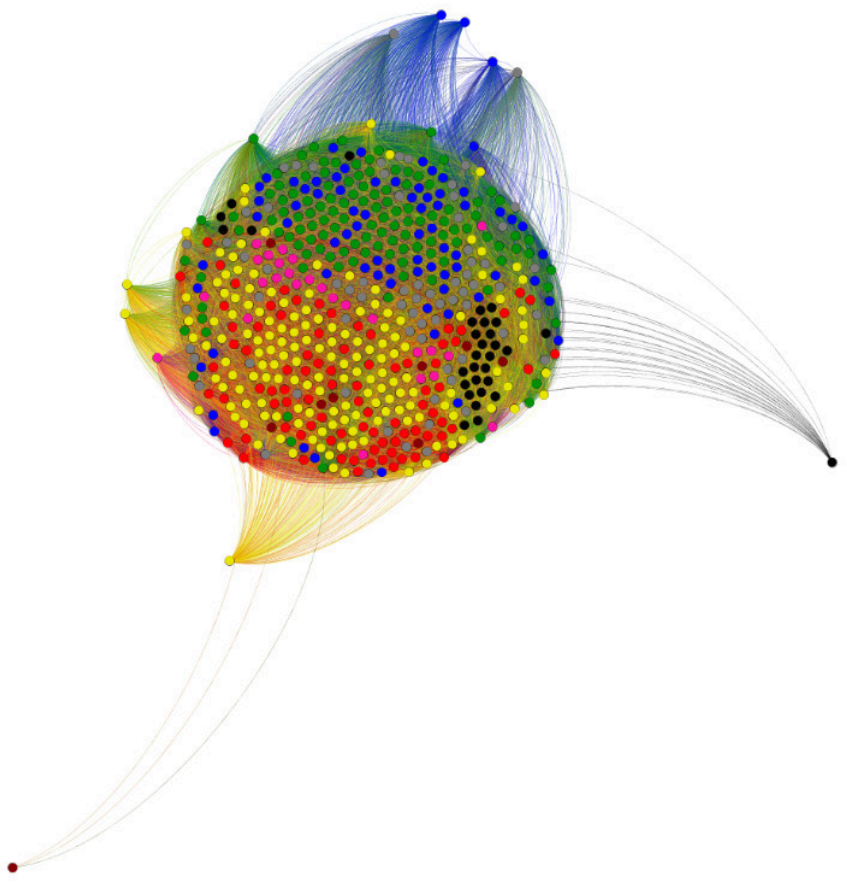

**(e)** Camera, *Draghi* cabinet.

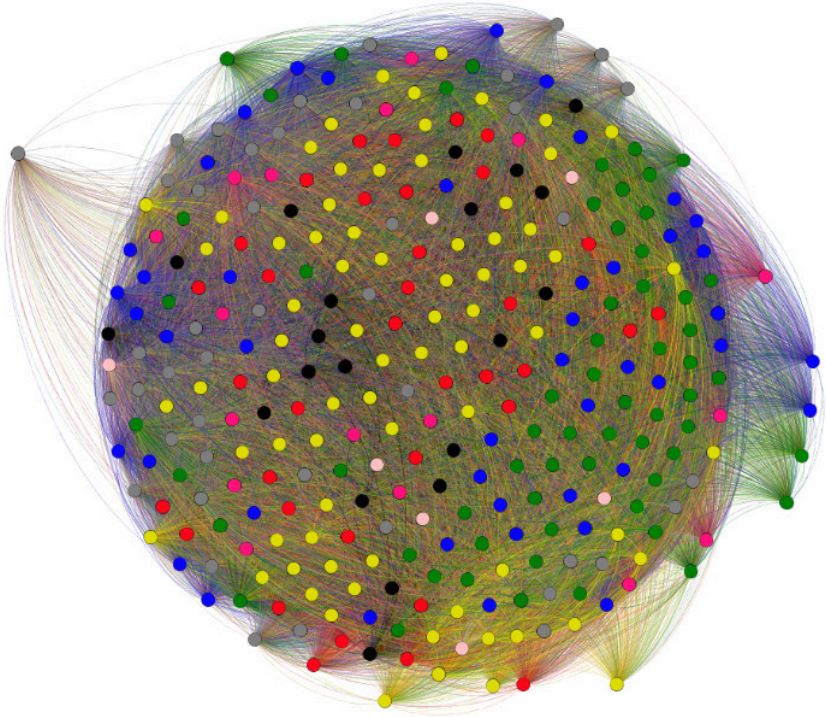

**(f)** Senato, *Draghi* cabinet.

**Figure 1. Co-voting networks in the first 50 days after cabinet formation, by cabinet and chamber.**
Panels show Conte I, Conte II, and Draghi governments for the Camera dei Deputati and the Senato della Repubblica (three cabinets × two chambers). Nodes represent Members of Parliament (MPs); edges represent co-voting similarity, with edge weight proportional to the fraction of identical votes. Node colours denote Parliamentary Groups; colour codes and acronyms are reported in Tables 0a – 0b.

Table 1 quantifies these structural differences through key network metrics. Network density remains high across all cabinets (94-99%), indicating strong overall interconnection among MPs – most representatives vote similarly on most bills, reflecting the legislative process where many votes are procedural or consensual. However, modularity values vary systematically with coalition type:

**Table 1. Structural Characteristics of Parliamentary Voting Networks**

| Cabinet | Chamber | Density | Modularity | Avg Degree Centrality | Avg Betweenness |
|---|---|---|---|---|---|
| Conte I | Camera | 98.5% | 0.27 | 0.99 | 0.000 |
| Conte I | Senato | 98.7% | 0.20 | 0.99 | 0.006 |
| Conte II | Camera | 97.4% | 0.10 | 0.97 | 0.000 |
| Conte II | Senato | 94.0% | 0.04 | 0.94 | 0.004 |
| Draghi | Camera | 95.3% | 0.08 | 0.95 | 0.002 |
| Draghi | Senato | 98.1% | 0.02 | 0.98 | 0.005 |

The **Conte I cabinet** (M5S-Lega coalition) shows the highest modularity (Q=0.27 Camera, 0.20 Senato), reflecting clear separation between the right-populist governing coalition and the center-left opposition. This partisan government produced the most structured voting divide.

The **Conte II cabinet** (M5S-PD-LEU-IV coalition) exhibits intermediate modularity (Q=0.10 Camera, 0.04 Senato) despite being a partisan coalition. The lower values possibly reflect the alliance between the – previously opponents, now forced into cooperation – populist M5S and center-left PD. Network density also drops to 94% in the Senato – the lowest across all cabinets – indicating increased voting fragmentation.

The **Draghi cabinet**, paradoxically, shows the *lowest* modularity (Q=0.08 Camera, 0.02 Senato) despite being a "grand coalition" including nearly all parties (M5S, Lega, FI, PD, IV, LEU) with only FDI in opposition. This counterintuitive finding reveals that broad formal support does not translate to unified voting behavior. Instead, parties maintain distinct identities through selective defection on specific bills, resulting in a network structure with less overall community separation than partisan governments.

Average degree centrality remains high (>0.94) across all networks, indicating most MPs are well-connected within the co-voting structure. The few nodes with low degree centrality correspond to MPs with distinctive voting patterns: ministers frequently absent for government duties, party-switchers in transition, or ideologically isolated independents (particularly within the heterogeneous Misto group).

Average betweenness centrality is near-zero for all networks, suggesting that most MPs are not structural bridges between communities – political divisions are sharp, with few mediators. We examine the small subset of high-betweenness nodes in Section 3.4.

## 3.2 Community Detection Exposes Implicit Coalitions

Applying modularity optimization to the co-voting networks reveals the actual voting blocs, which can differ substantially from declared political alliances. Figure 2 displays the distribution of Political Groups across detected communities for each cabinet and chamber.

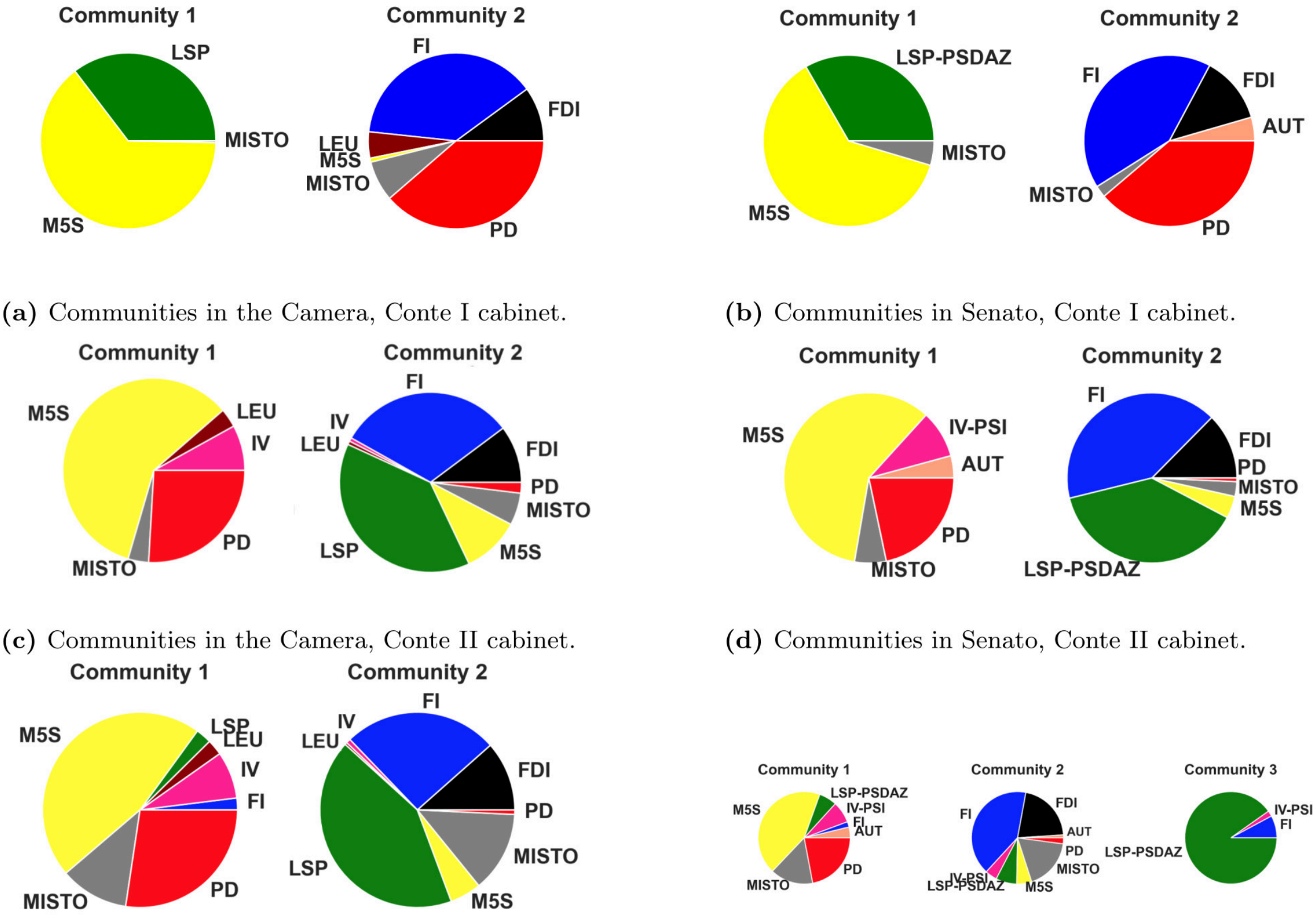

(a) Communities in the Camera, Conte I cabinet.

(b) Communities in Senato, Conte I cabinet.

(c) Communities in the Camera, Conte II cabinet.

(d) Communities in Senato, Conte II cabinet.

(e) Communities in Camera, Draghi cabinet.

(f) Communities in Senato, Draghi cabinet.

**Figure 2. Political Group composition of detected communities across cabinets and chambers.**
For each cabinet and chamber, pie charts show the fraction of Members of Parliament from each Parliamentary Group within each detected community. Communities are identified via modularity maximisation on the co-voting networks shown in Figure 2. Pie chart areas are not proportional to community size. Parliamentary Group colours and acronyms are defined in Tables 0a – 0b.

**Table 2. Community Sizes vs. Formal Coalitions** Note that for the Draghi cabinet, formal majority does not correspond to a single government-aligned community (see Figure 3e-f).

| Cabinet | Chamber | Comm. 1 | Majority | Comm. 2 | Opposition | Comm. 3 |
|---|---|---|---|---|---|---|
| Conte I | Camera | 325 | 326 | 269 | 268 | |
| Conte I | Senato | 174 | 166 | 134 | 142 | |
| Conte II | Camera | 301 | 343 | 311 | 269 | |
| Conte II | Senato | 166 | 173 | 143 | 136 | |
| Draghi | Camera | 307 | 500 | 269 | 76 | |
| Draghi | Senato | 159 | 268 | 95 | 37 | 51 |

**Conte II Cabinet** (Figure 3c-d): Community structure reveals significant defection from the formal coalition. In the Camera, 15% of M5S members and 7% of PD members appear in Community 2 (opposition-aligned), resulting in Community 2 (311 MPs) being *larger* than Community 1 (301 MPs) despite Community 1 representing the formal majority (343 MPs). This numerical inversion demonstrates that declared government support did not translate to consistent voting behavior – a substantial faction voted more frequently with the opposition. The Senato shows similar patterns but less pronounced.

This finding explains the lower modularity for Conte II: while two communities exist, their composition is muddled, with substantial cross-over between formal and actual coalitions. The fragile alliance between M5S and PD is evident in voting behavior even during the coalition's honeymoon period.

**Draghi Cabinet (Figure 3e-f).** The technical government produces the most complex community structure.
In the Camera (Figure 3e), the community most closely aligned with government voting is **Community 2**, dominated by Lega Salvini Premier and Forza Italia. In contrast, **Community 1** aggregates members of Movimento 5 Stelle, Partito Democratico, and the Misto group, displaying a markedly heterogeneous voting pattern. This asymmetry indicates that, despite broad formal support, only a subset of governing parties forms a structurally coherent voting bloc, while the remainder exhibit fragmented or conditional alignment.

The Senato (Figure 3f) shows an even more pronounced fragmentation, with the emergence of a **third community** composed primarily of LSP-PSDAZ members (73% of the group), together with Italia Viva-PSI. This community is structurally distinct from both the government-aligned Community 2 and the opposition-aligned Community 1, representing a form of **selective support** that cooperates on specific issues while diverging on others.

The discrepancy between the size of the government-aligned communities (Community 2 in the Camera; Communities 2 and 3 in the Senato) and the formal parliamentary majority is substantial. In the Camera, the structurally coherent voting bloc represents less than half of the declared majority. This quantitatively confirms that the Draghi government relied on fragmented and issue-dependent support rather than on a unified voting coalition.

## 3.3 Party Coherence Dynamics Track Political Transitions

We now examine the evolution of the intra-group coherence to understand how party discipline changes across government formations. Figure 3 displays coherence trajectories for each Political Group.

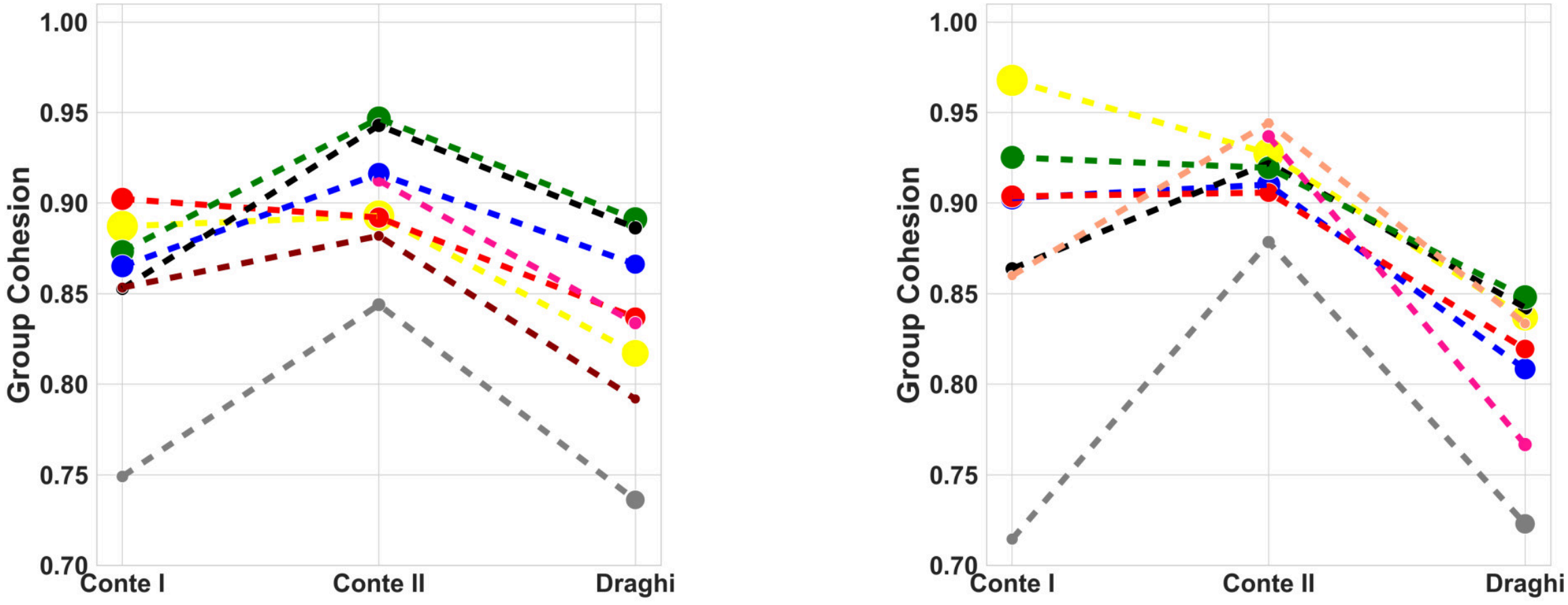

(a) Coherence evolution within Camera groups.

(b) Coherence evolution within Senato groups.

**Figure 3. Evolution of intra-group voting coherence across cabinets, by chamber.** For each Parliamentary Group, points report the intra-group coherence measured over the post-cabinet formation window (first 50 days) for each government, shown separately for the Camera dei Deputati and the Senato della Repubblica. Marker size is proportional to group size. Error bars indicate the standard deviation of individual members' coherence within each group. Parliamentary Group colours and acronyms are defined in Tables 0a – 0b.

Several patterns emerge:

**The Misto group** consistently shows the lowest coherence (0.73-0.78), as expected for a heterogeneous collection of independents, small parties, and recent defectors lacking unified political program. This validates our coherence metric – it correctly identifies the least disciplined group.

**Governing to opposition transitions reduce coherence**: In the Senato (Figure 4b), Lega and M5S both exhibit sharp coherence drops when transitioning from Conte I (governing) to Conte II (opposition for Lega) and continuing decline through Draghi (opposition/ambiguous). Lega drops from 0.90 (Conte I) to 0.88 (Conte II) to 0.85 (Draghi). M5S follows similar trajectory: 0.97 (Conte I) → 0.93 (Conte II) → 0.85 (Draghi). This suggests that opposition status – or possibly the ambiguous positioning in Draghi's case – weakens party discipline, as members have more freedom to defect without threatening government stability.

**PD coherence decline in Camera**: The center-left PD party shows declining coherence in the Camera across all three governments (0.90 → 0.88 → 0.85), despite being in opposition (Conte I), government (Conte II), and government (Draghi). This reflects internal tensions between the party's traditional center-left base and those willing to ally with populists (M5S) or support technocrats (Draghi). The community analysis (Section 3.2) revealed that 7% of PD members voted with opposition in Conte II, confirming this internal division.

**All groups lose coherence under Draghi**: Every Political Group except FDI (the sole opposition party) shows coherence decline from Conte II to Draghi. This universal pattern reflects the ambiguous nature of the technical government – parties formally support Draghi but struggle to maintain discipline when voting on specific policies that conflict with ideological positions. The framework captures this "support without unity" phenomenon quantitatively.

**FDI maintains high coherence**: Fratelli d'Italia, the only party in opposition throughout all three cabinets, maintains consistently high coherence (0.88-0.90). This suggests that clear opposition positioning – without the complications of government participation – facilitates party unity.

Figure 4 complements this analysis by examining individual member loyalty. The boxplots show that most MPs have high loyalty to their groups (median >0.85), but outliers exist in every party. These outliers are candidates for party switching or represent ideologically distinct factions. For instance, in the Draghi Camera (Figure 5e), several M5S, Lega, and FI members show loyalty <0.70 despite formal group membership, quantifying the defection patterns identified in community detection.

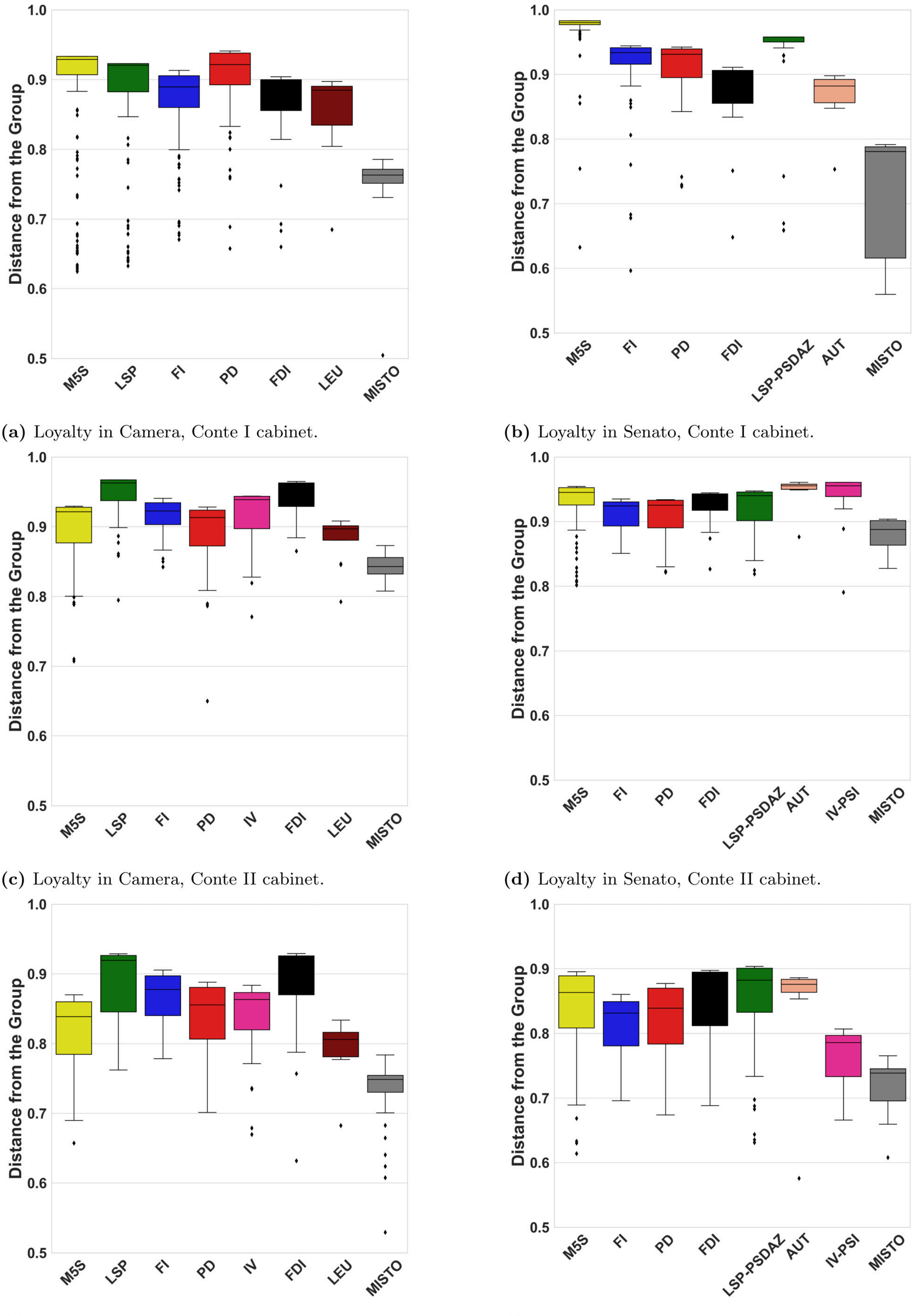

(a) Loyalty in Camera, Conte I cabinet.

(b) Loyalty in Senato, Conte I cabinet.

(c) Loyalty in Camera, Conte II cabinet.

(d) Loyalty in Senato, Conte II cabinet.

(e) Loyalty in Camera, Draghi cabinet.

(f) Loyalty in Senato, Draghi cabinet.

**Figure 4. Distribution of individual Parliamentary Group loyalty across cabinets and chambers.**
For each cabinet and chamber, boxplots report the distribution of individual Members of Parliament loyalty (defined as the coefficient of location within their Parliamentary Group). Boxes indicate the median and interquartile range, with whiskers extending to non-outlier values; dots denote outliers corresponding to individual Members of Parliament with unusually low or high loyalty. Panels correspond to the Conte I, Conte II, and Draghi governments for the Camera dei Deputati and the Senato della Repubblica.

## 3.4 Betweenness Centrality Identifies Concentrated Mediator Power

While most MPs vote with clear partisan blocs, a small subset acts as bridges between communities. Betweenness centrality (Equation 3) quantifies this structural position. Across all six networks (3 cabinets × 2 chambers), only 21 MPs (2.1% of parliament) exhibit betweenness significantly greater than zero (>0.1 on normalized scale).

These high-betweenness MPs fall into distinct categories:

1. **Party-switchers**: MPs who changed parliamentary groups during the legislature or were about to switch within weeks of the analyzed period. Their transitional status places them structurally between their former and future groups.

2. **Veteran politicians with cross-party relationships**: Long-serving MPs with established relationships across party lines, enabling them to coordinate votes on specific issues despite formal opposition.

3. **Ideologically moderate members**: Representatives from parties' centrist wings who occasionally vote with the other bloc on particular policies.

The concentration of betweenness is striking: the top 2% of MPs account for essentially all bridge-building capacity in the network. This suggests that mediator power in polarized parliaments is highly concentrated, not distributed across the membership. Formal party leadership (ministers, group leaders) does not always overlap with high betweenness – some mediators operate informally.

**The Immigration Case Study Anomaly**: In the 2018 immigration votes (Conte I, Camera – see Section 3.5), several deputies show unusually high betweenness (0.3-0.8 range). These are primarily members who would later switch from M5S to Misto, suggesting that their voting behavior on this polarizing topic signaled ideological distance from their party, presaging eventual defection. The framework thus provides early indicators of party instability.

Conversely, in the 2020 immigration vote network (Conte II, Senato), the graph is nearly disconnected – two tight clusters with minimal bridges – resulting in zero or near-zero betweenness for all senators. This reflects complete polarization: no mediators existed on this issue during this period.

The average betweenness values in Table 1 (all <0.01) confirm that, in the period analysed, bridge-building is exceptional, not typical. Parliamentary politics was characterized by sharp divides with rare mediators, rather than continuous spectrum with many intermediates.

## 3.5 Case Study: Immigration Votes Validate Framework on Polarizing Topics

To test whether the framework captures dynamics beyond general legislative voting, we analyze roll-call votes specifically on immigration policy – a highly polarizing topic in Italian politics (Di Mauro and Verzichelli, 2019; Schmidt et al., 2020; Urso, 2018). We examine two sets of immigration-related bills: one during Conte I (October-November 2018, Camera and Senato) and one during Conte II (October-December 2020, Camera and Senato).

Figure 5 displays the co-voting networks for these immigration votes, with Figure 7 showing the corresponding community structures.

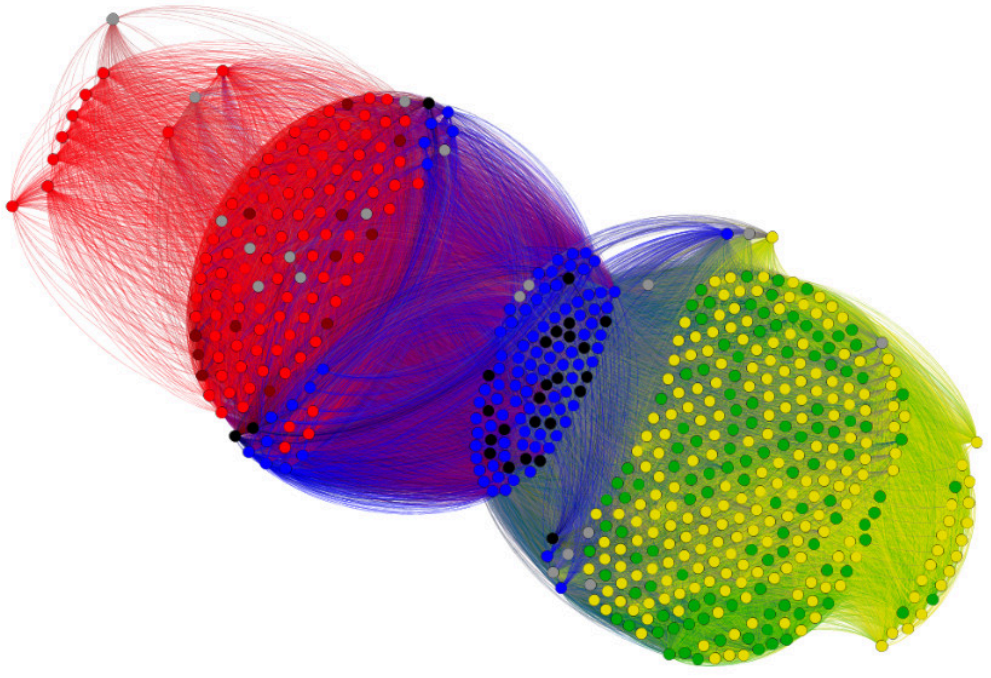

**(a)** Network structure of votes in Camera in the following dates: 2018-11-14, 2018-11-26/27/28.

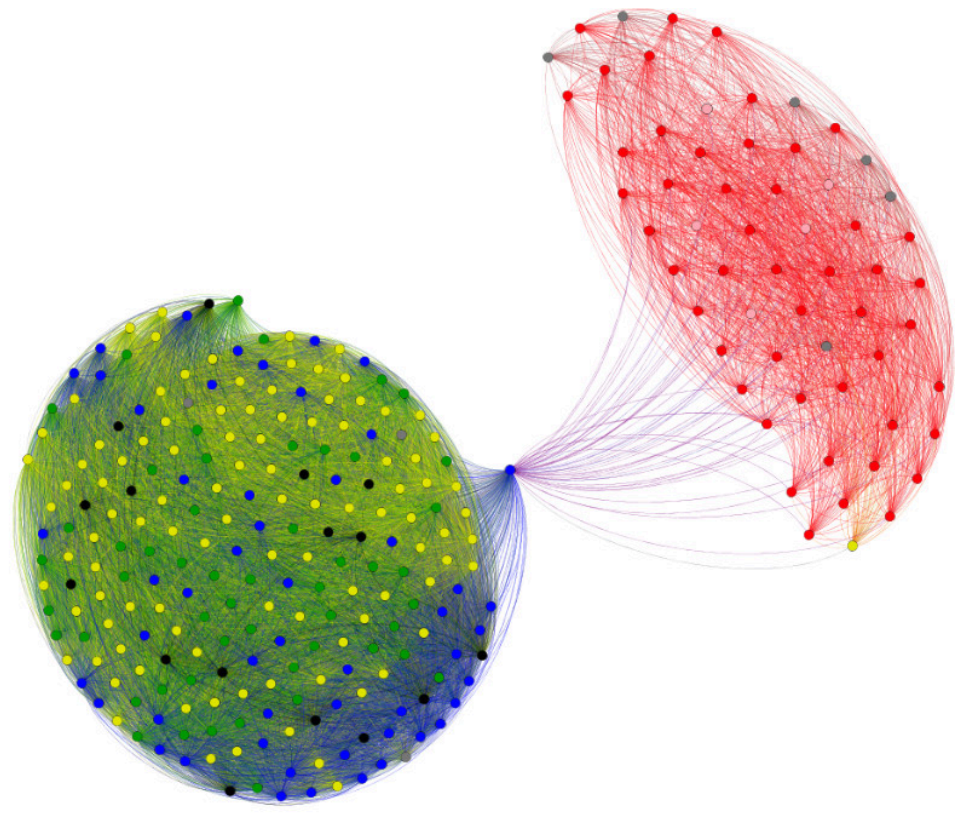

**(b)** Network structure of votes in Senato in the following dates: 2018-10-16, 2018-11-06.

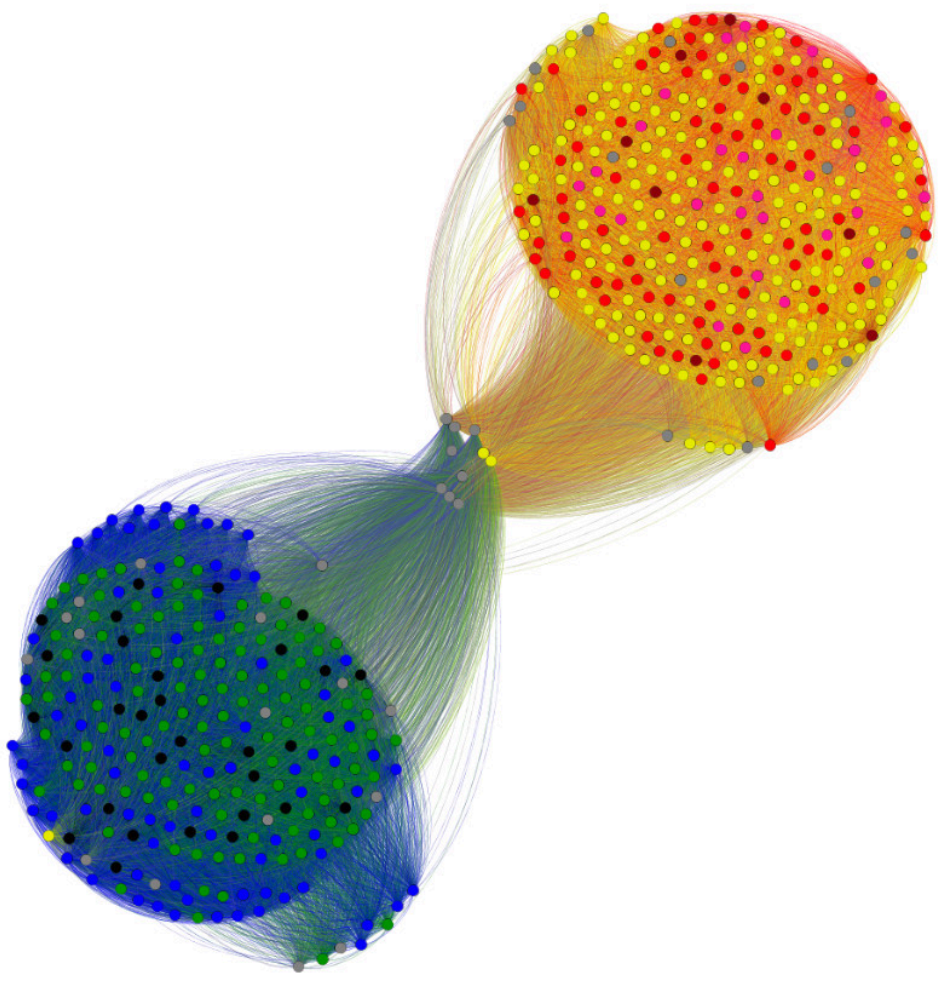

**(c)** Network structure of votes in Camera in the following dates: 2020-10-28, 2020-11-30, 2020-12-09.

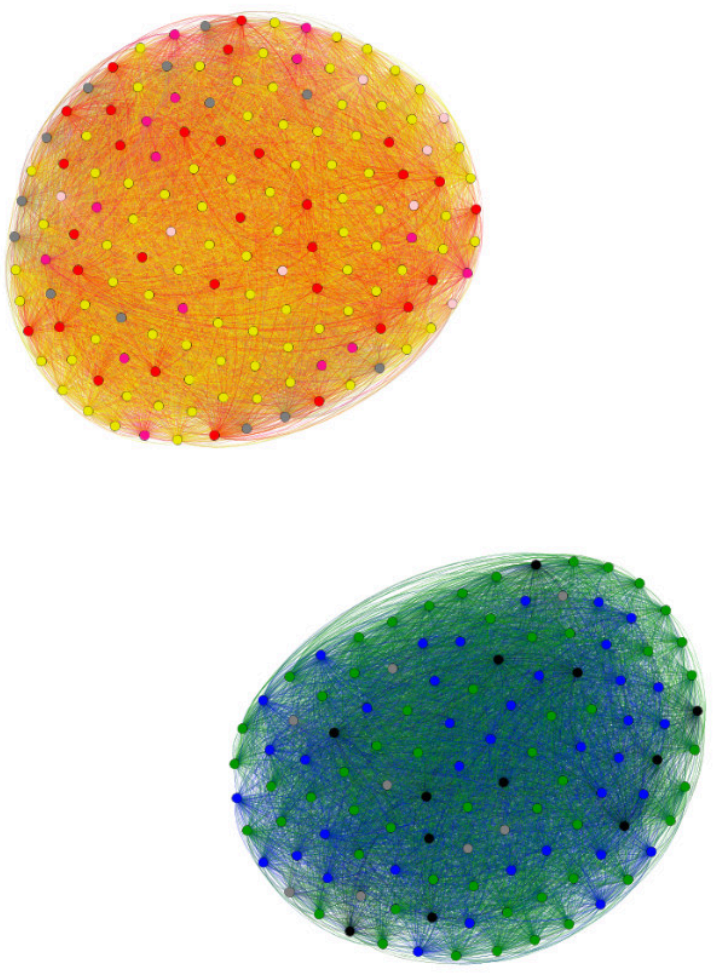

**(d)** Network structure of votes in Senato in the following date: 2020-12-17.

**Figure 5 Co-voting networks for immigration-related roll-call votes, by cabinet and chamber.**

Panels (a – b) show immigration votes during the Conte I government (October – November 2018) for the Camera dei Deputati and the Senato della Repubblica; panels (c – d) show immigration votes during the Conte II government (October – December 2020) for the Camera dei Deputati and the Senato della Repubblica. Nodes represent Members of Parliament; edges represent co-voting similarity, with edge opacity proportional to the frequency of identical votes. Node colours denote Parliamentary Groups; colour codes and acronyms are reported in Tables 0a – 0b.

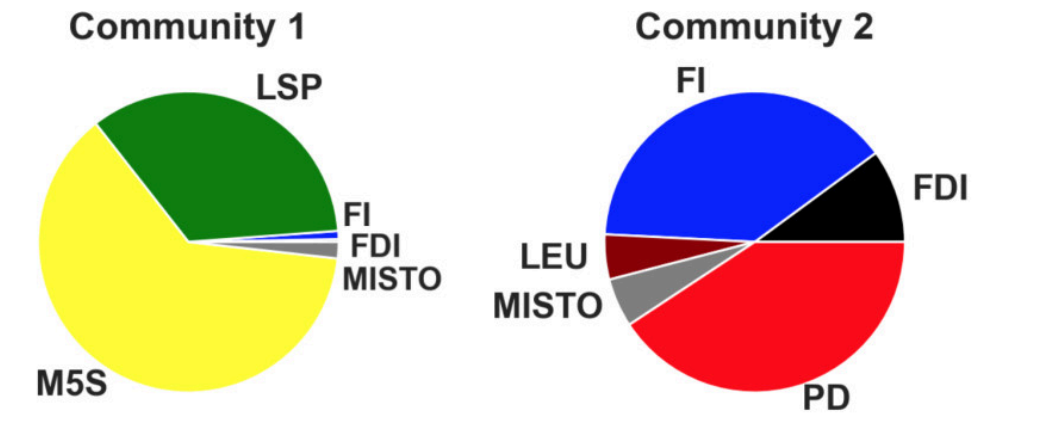

(a) Community structure of votes in Camera in the following dates: 2018-11-14, 2018-11-26/27/28.

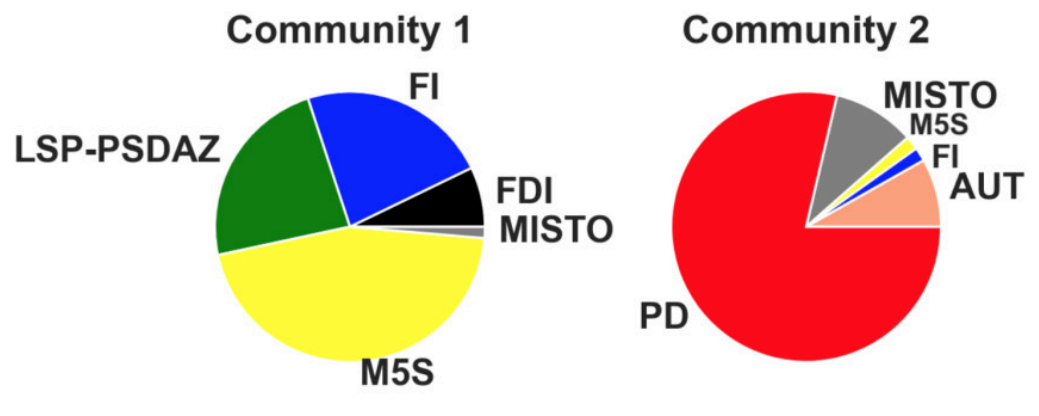

(b) Community structure of votes in Senato in the following dates: 2018-10-16, 2018-11-06.

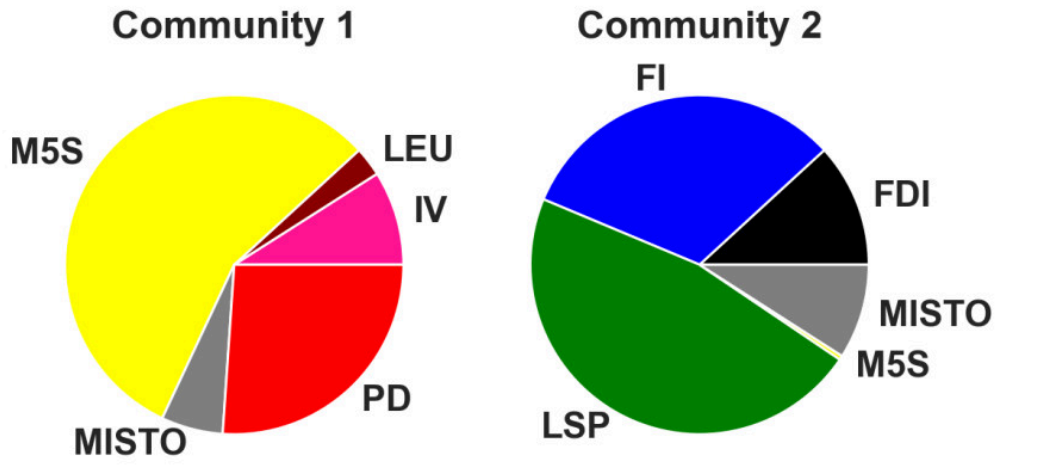

(c) Community structure of votes in Camera in the following dates: 2020-10-28, 2020-11-30, 2020-12-09.

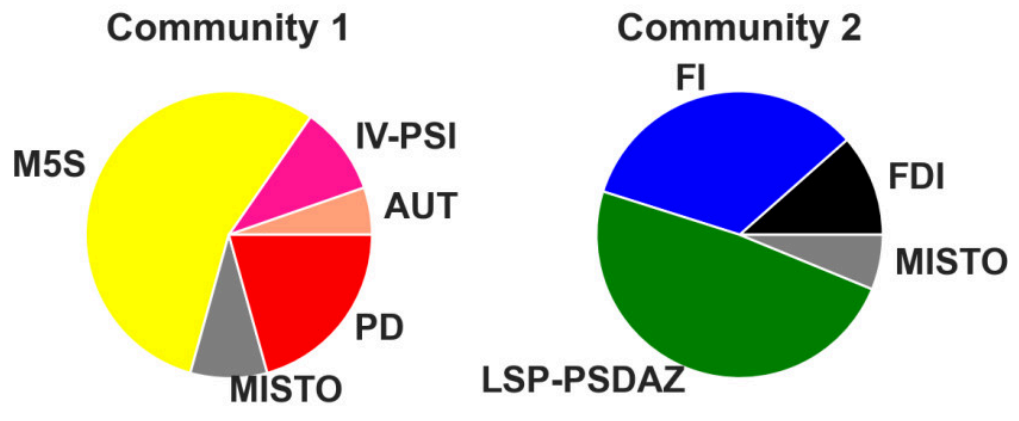

(d) Community structure of votes in Senato in the following date: 2020-12-17.

**Figure 6. Community structure of immigration-related co-voting networks.** Same networks and panel structure as in Figure 6, with nodes coloured according to the communities detected via modularity maximisation. Node colours within communities indicate Parliamentary Groups, following the colour codes and acronyms reported in Tables 0a – 0b. Results are shown separately for the Camera dei Deputati and the Senato della Repubblica and for the Conte I and Conte II governments.

**Conte I Immigration Votes** (2018): The networks show clear bipartite structure – two dense lobes with sparse connections – matching expectations for a polarizing issue. The governing coalition (M5S-Lega) voted cohesively for restrictive immigration policies, while opposition parties (PD, FI, LEU) voted against. The Senato network (Figure 6b) is nearly disconnected, with only one senator showing significant betweenness (0.6) – a single bridge between communities. This senator was a veteran politician with cross-party relationships who attempted to mediate on specific amendments.

**Conte II Immigration Votes** (2020): The 2020 networks show similarly clear separation, but now the coalitions have inverted. M5S, having switched from alliance with right-wing Lega to alliance with center-left PD, now votes for more liberal immigration policies. Lega and FI, formerly government partners of M5S, now vote in opposition. This 180-degree reversal demonstrates that parties' positions on immigration are conditional on coalition membership, not fixed ideological commitments – a finding enabled by longitudinal network analysis.

The Camera 2020 network (Figure 6c) shows seven deputies with elevated betweenness (0.1-0.4 range): five had recently switched or were about to switch from M5S to Misto, while two were long-standing Misto members. These mediators represent the personal cost of M5S's coalition pivot – members ideologically committed to immigration restriction could not follow the party's policy reversal.

**The Modularity Paradox**: Table 3 compares modularity for immigration votes versus general votes:

**Table 3. Modularity Comparison: General vs. Immigration Votes**

| Cabinet | Chamber | General Votes (Q) | Immigration Votes (Q) |
|---|---|---|---|
| Conte I | Camera | 0.27 | 0.21 |
| Conte I | Senato | 0.20 | 0.10 |
| Conte II | Camera | 0.10 | 0.45 |
| Conte II | Senato | 0.04 | 0.46 |

Surprisingly, **Conte I shows *lower* modularity on immigration than on general votes**, despite immigration being a coalition-defining issue for the M5S-Lega government. How can a polarizing topic produce less network separation?

The explanation lies in opposition structure. During Conte I, the opposition was fragmented: PD, FI, FDI, and LEU often disagreed on amendments and specific provisions, even while all voting against the overall restrictive direction. Some FI members (center-right, traditionally pro-business) occasionally voted with the government on specific economic provisions related to immigration. This opposition fragmentation reduces overall modularity even though the government was cohesive.

Conversely, **Conte II shows *higher* modularity on immigration (Q=0.45-0.46) than on general votes (Q=0.10-0.04)**. The opposition to M5S-PD's more liberal immigration policies was united and disciplined – Lega, FI, and FDI voted as a cohesive bloc. The increased modularity reflects this unified opposition, not increased government cohesion.

This finding has important methodological implications: **modularity captures system-wide polarization structure, not just government-opposition separation**. A cohesive government facing fragmented opposition produces lower modularity than a shaky government facing unified opposition. Traditional roll-call analysis, focusing only on government vote shares, would miss this distinction.

The immigration case study thus validates the framework's ability to reveal nuanced dynamics on specific issues while demonstrating that interpretation requires considering both majority and opposition structure – precisely what network analysis enables by treating all pairwise relationships simultaneously.

# 4 Discussion

Across multiple analytical scales, the proposed framework reveals structural patterns that, in the context of the Italian Parliament during the 18th legislature, remain largely invisible to traditional legislative analysis.

First, coalition type is associated with systematic differences in network modularity. In the Italian case examined here, narrow partisan coalitions, such as the Conte I government, display higher levels of separation, whereas technical governments, most notably the Draghi cabinet, exhibit lower modularity despite broader formal parliamentary support. This suggests that, in this setting, coalition breadth does not necessarily translate into voting cohesion and that system-level polarization is not determined by majority size alone.

Second, declared political alliances frequently diverge from observed voting behaviour. During the Conte II cabinet, the nominal governing majority included substantial factions that regularly voted with the opposition. Under the Draghi government, the formal grand coalition fragmented into multiple voting blocs in the Senato. These patterns indicate that voting alignment emerges endogenously from interaction structures rather than being fully determined by formal agreements.

Third, periods of political transition are associated with weakened party discipline. Groups moving between government and opposition roles, such as Lega and Movimento 5 Stelle, experience notable reductions in internal coherence. Under the technical government, all parties except the sole opposition group (Fratelli d'Italia) display lower coherence, suggesting that ambiguity in political positioning may undermine collective discipline in similar institutional contexts.

Fourth, mediator power within the parliamentary voting network appears highly concentrated. In this case, only about 2% of Members of Parliament act as structural bridges between communities. These actors are often party switchers or experienced legislators with cross-group ties, indicating that brokerage reflects strategic positioning within the network rather than formal leadership roles.

Finally, polarization in issue-specific voting networks depends on the structure of the system as a whole rather than on government cohesion alone. In immigration-related votes, opposition fragmentation during the Conte I cabinet is associated with reduced modularity despite a cohesive governing coalition, whereas a more unified opposition during the Conte II cabinet corresponds to higher polarization. This highlights opposition structure as a key factor shaping system-level polarization in this empirical setting.

Taken together, these findings indicate that the framework captures meaningful structural regularities across coalition types, issue domains, and temporal phases within the Italian parliamentary context. While these patterns should not be interpreted as universal laws, they illustrate how network-based analysis can reveal relational mechanisms that may be relevant in other multi-party legislatures and political settings.

# 5 Conclusions

This work introduces a general network-science framework for analysing collective decision-making in legislatures, integrating standard network measures to capture political structure at the individual, group, and system levels.

Applied to over four million roll-call votes from the Italian Parliament during a period of exceptional political turbulence (2018-2021), the framework reveals patterns that escape traditional approaches to legislative analysis. In particular, it shows that formal coalition size is a poor predictor of voting cohesion, that polarization emerges from system-wide interaction structure rather than declared alliances, and that individual brokerage power is highly concentrated.

Although validated on the Italian case, the framework is not context-specific. It relies exclusively on public roll-call data and requires no institutional parameters, making it directly applicable to any legislature with transparent voting records. The increasing availability of machine-readable parliamentary data through open government initiatives further enhances its relevance.

More broadly, an important implication of these results is that generalisability itself becomes a substantive theoretical and comparative research question. Determining whether the structural patterns observed here reflect broader features of multi-party legislatures or instead arise from specific institutional and political conditions requires systematic cross-case comparison. Network-based approaches are particularly well suited to this task, as they provide a common analytical language for comparing legislative systems across countries, time periods, and issue domains without relying on case-specific institutional assumptions.

Building on this perspective, the framework opens several concrete directions for future research. Cross-national applications could assess whether similar structural regularities emerge in different institutional settings. Longitudinal analyses could investigate whether network metrics anticipate coalition breakdown or government survival over time. Issue-specific network studies could examine whether parties maintain stable alliances across policy domains or instead form context-dependent coalitions.

Beyond academic research, the methodology provides quantitative infrastructure for democratic accountability. Real-time implementations could enable journalists, civic organisations, and citizens to verify whether representatives vote consistently with their declared positions, assess party discipline, and identify key intermediaries in legislative decision-making.

By treating voting behaviour as a relational system rather than as a sequence of aggregate vote counts or party-level metrics, this work positions network analysis as analytical infrastructure for the systematic study of political and organisational systems. As demands for transparency and evidence-based accountability grow, network-based frameworks of this kind offer powerful tools for understanding and monitoring collective political behaviour.

# Data Availability

All data used in this study are publicly available through the Italian Parliament's open data SPARQL endpoints: http://dati.camera.it/sparql (Camera dei Deputati) and https://dati.senato.it/sito/23 (Senato della Repubblica). The complete database schema, data collection scripts, analysis code, and documentation will be made available at [GitHub repository URL] upon publication. This ensures full reproducibility and facilitates adaptation to other parliamentary systems.

# Code Availability

All analysis was performed using Python 3.9 with standard scientific computing libraries (NumPy, Pandas, NetworkX, SQLAlchemy). Complete code for network construction, community detection, coherence calculations, and visualization will be released as open-source software with documentation under MIT license at [https://github.com/francescacollu/VotingBehavior].

# Acknowledgments

The authors received no specific funding for this work. We thank colleagues for helpful discussions and comments on earlier versions of the manuscript.

# Author Contributions

FC: Data collection, database design, network analysis, writing. AS: Conceptualization, methodology, writing, supervision. ELN: Visualization, coherence analysis, writing. All authors reviewed and approved the final manuscript.

# Competing Interests

The authors declare no competing interests.

# Appendix A: Technical Details on Data Collection and Database Structure

This appendix provides comprehensive technical information on data collection procedures, database schema, and implementation details that supplement the main text's methodological overview.

## A.1 Data Sources and Access Methods

The Italian Parliament provides open data access through two SPARQL endpoints:

- **Camera dei Deputati** (Lower House): http://dati.camera.it/sparql
- **Senato della Repubblica** (Upper House): https://dati.senato.it/sito/23

Both endpoints implement the SPARQL Protocol and RDF Query Language, enabling structured queries against semantic web data. However, each chamber's endpoint has different characteristics and limitations requiring adapted collection strategies.

### A.1.1 Camera Data Collection

The Camera SPARQL endpoint allows downloading query results in CSV format but imposes a hard limit of 10,000 records per query. Since individual deputies cast thousands of votes over the legislature, direct queries for all votes by all deputies exceed this limit.

Our workaround strategy:

1. Execute SPARQL query to retrieve the URL pattern for accessing individual deputy vote records
2. Construct URL requests for each deputy ID over limited time periods (typically one year)
3. Iterate through all deputy IDs and time periods systematically
4. Download and aggregate CSV files containing votes for every Camera member

This approach required executing approximately 3,200 individual requests (632 deputies × ~5 time periods) over several days to avoid overloading the server.

**Example SPARQL query for Camera data:**

```
PREFIX ocd: <http://dati.camera.it/ocd/>
PREFIX dc: <http://purl.org/dc/elements/1.1/>
PREFIX rdfs: <http://www.w3.org/2000/01/rdf-schema#>

SELECT ?votazione ?data ?voto ?deputato ?nome
WHERE {
  ?votazione ocd:rif_deputato ?deputato .
  ?votazione ocd:voto ?voto .
  ?votazione dc:date ?data .
  ?deputato rdfs:label ?nome .
  FILTER (?data >= "2018-03-23"^^xsd:date && ?data <= "2019-03-
23"^^xsd:date)
  FILTER (?deputato = ocd:deputato.XXXXX)
```

```
}
ORDER BY ?data
```

Where `XXXXX` is replaced with each deputy's unique identifier.

### A.1.2 Senato Data Collection

The Senato website does not enforce the 10,000 record limit on its data portal but lacks a unified SPARQL endpoint for all vote data. Instead, vote information is accessible through a web-based search form at a dedicated section of the website.

Our collection strategy:

1. Use Selenium WebDriver (Python library for browser automation) to interact with the search form programmatically
2. Submit queries for each senator across different time periods
3. Parse HTML results and extract vote data
4. Structure extracted data into consistent format matching Camera data

This web scraping approach required approximately 315 automated browser sessions (315 senators) and was more fragile than the Camera's structured SPARQL approach, requiring error handling for timeout issues, page structure changes, and intermittent server availability.

**Selenium automation pseudocode:**

```python
from selenium import webdriver
from selenium.webdriver.common.by import By

driver = webdriver.Firefox()
driver.get("https://dati.senato.it/votazioni")

# For each senator
for senator_id in senator_ids:
    # Fill search form
    driver.find_element(By.ID, "senator_select").send_keys(senator_id)
    driver.find_element(By.ID, "date_from").send_keys("23/03/2018")
    driver.find_element(By.ID, "date_to").send_keys("23/09/2021")
    driver.find_element(By.ID, "submit_button").click()

    # Parse results table
    results = driver.find_elements(By.CLASS_NAME, "vote_row")
    for row in results:
        extract_vote_data(row)
```

## A.2 Database Schema and Organization

We organized collected data into a relational database using MySQL with SQLAlchemy (Bayer, 2012) as the Python ORM (Object-Relational Mapping) layer. Figure A1 illustrates the complete schema.

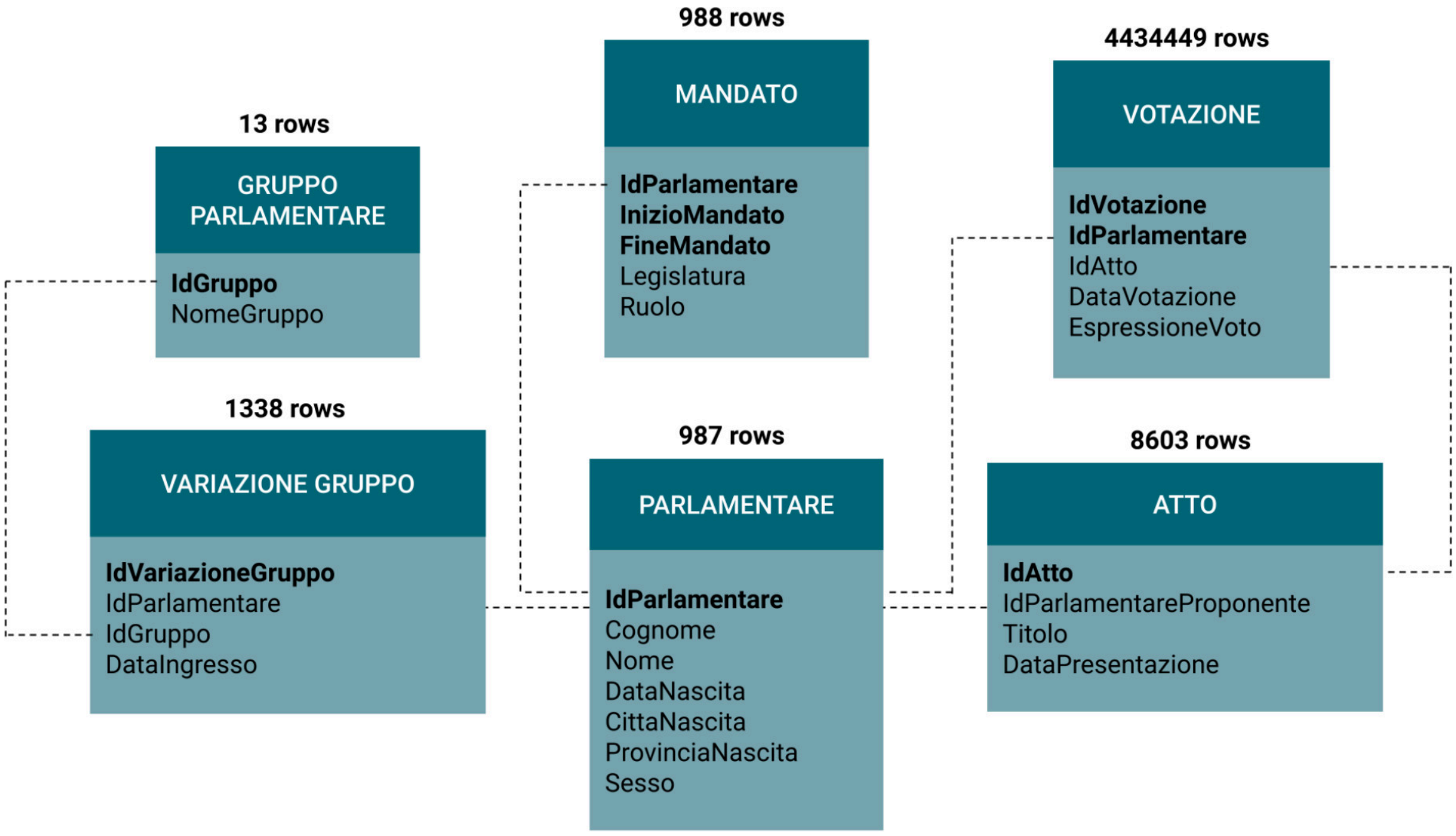

**Figure A1. Relational database schema for vote collection and network construction.** Entity – relationship diagram of the relational database used in this study. Field names in bold denote primary keys; dashed connectors link primary keys to foreign keys in related tables. Numbers above tables indicate row counts. The schema supports tracking parliamentarians, mandates, parliamentary groups, group changes (party switching), legislative acts, and individual roll-call vote expressions, enabling reconstruction of time-resolved MP – group affiliations and vote matrices for network analysis.

The schema consists of six main tables:

## Table: PARLAMENTARE (987 rows)

Stores individual parliamentarian information:

- **IdParlamentare** (Primary Key): Unique identifier for each MP
- Cognome: Last name
- Nome: First name
- DataNascita: Date of birth
- CittaNascita: City of birth
- ProvinciaNascita: Province of birth
- Sesso: Gender

*Note:* 987 unique parliamentarians across 988 mandates because one deputy was elected to Camera then moved to Senato during the same legislature – the first such occurrence in Italian parliamentary history.

## Table: MANDATO (988 rows)

Tracks parliamentary terms and chamber assignments:

- **IdMandato** (Primary Key): Unique mandate identifier
- IdParlamentare (Foreign Key → PARLAMENTARE): Links to parliamentarian
- InizioMandato: Mandate start date
- FineMandato: Mandate end date (NULL if still active)
- Legislatura: Legislature number (18 for all records)
- Ruolo: Role (Deputato for Camera, Senatore for Senato)

### Table: GRUPPO_PARLAMENTARE (13 rows)

Defines parliamentary groups:

- **IdGruppo** (Primary Key): Unique group identifier
- NomeGruppo: Full group name

Includes 8 Camera groups and 8 Senato groups (some overlap in naming but different IDs).

### Table: VARIAZIONE_GRUPPO (1,338 rows)

Tracks party switching and group membership changes:

- **IdVariazioneGruppo** (Primary Key): Unique change identifier
- IdParlamentare (Foreign Key → PARLAMENTARE): Links to MP
- IdGruppo (Foreign Key → GRUPPO_PARLAMENTARE): Links to new group
- DataIngresso: Date of joining group

Enables temporal tracking of party affiliations, critical for accurate network construction at any time point.

### Table: ATTO (8,603 rows)

Contains information on bills and legislative acts:

- **IdAtto** (Primary Key): Unique bill identifier
- IdParlamentareProponente: Proposing MP (if applicable)
- Titolo: Bill title
- DataPresentazione: Submission date

### Table: VOTAZIONE (4,434,449 rows)

The core table storing individual vote expressions:

- **IdVotazione** (Primary Key): Unique vote record identifier
- IdParlamentare (Foreign Key → PARLAMENTARE): Links to voting MP
- IdAtto (Foreign Key → ATTO): Links to bill being voted on
- DataVotazione: Date and time of vote
- EspressioneVoto: Vote expression (Favorevole/Contrario/Astenuto/Assente)

This table's 4.4 million records constitute the primary analytical data, with each row representing one MP's vote on one bill.

## A.3 Data Quality and Preprocessing

Several data quality issues required preprocessing:

**Missing values**: Approximately 0.3% of vote records lacked MP identifiers due to data entry errors or system glitches. These records were excluded from analysis.

**Inconsistent date formats**: Camera and Senato used different datetime formats requiring standardization to ISO 8601 format (YYYY-MM-DD HH:MM:SS).

**Vote expression variations**: Different spellings and abbreviations for vote types (e.g., “Favorevole” vs “Fav” vs “F”) required normalization to standard categories: {Favorevole, Contrario, Astenuto, Assente}.

**Party switching temporal precision**: Group membership changes are dated but not time-stamped. We assign group membership at the date level, assuming switches occur at start of day. For votes occurring on the same day as a group switch, we assign the MP to their *new* group, as switches typically occur before the day’s voting session.

**Duplicate records**: Approximately 150 duplicate vote records (same MP, same bill, same date) were identified and removed, retaining the first occurrence.

## A.4 Parliamentary Group Compositions

Tables A1 and A2 provide detailed information on parliamentary group sizes and compositions at the time of each cabinet’s investiture vote.

**Table A1. Distribution of seats in the Camera dei Deputati across the three cabinets of the 18th legislature.**
Seat counts and percentages refer to the composition of the Camera dei Deputati at the time of each cabinet’s investiture vote. Percentages are computed relative to the total number of Camera seats (630), excluding temporary vacancies. Italia Viva (IV) was founded during the Conte II cabinet (September 2019) following a split from the Partito Democratico.

| Group | Conte I (June 2018) | Conte II (September 2019) | Draghi (February 2021) |
|---|---|---|---|
| M5S | 222 (35.3%) | 215 (34.2%) | 169 (26.8%) |
| LSP | 125 (19.9%) | 125 (19.9%) | 133 (21.1%) |
| FI | 105 (16.6%) | 98 (15.6%) | 86 (13.6%) |
| PD | 111 (17.7%) | 113 (17.9%) | 96 (15.3%) |
| IV | – | – | 30 (4.7%) |
| FDI | 30 (4.8%) | 34 (5.4%) | 33 (5.2%) |
| LEU | 14 (2.2%) | 14 (2.3%) | 13 (2.0%) |
| MISTO | 21 (3.4%) | 30 (4.8%) | 54 (8.6%) |
| **Majority** | **347** | **343** | **500** |
| **Opposition** | **268** | **269** | **76** |

Note: Seats may not sum to 630 due to vacancies. The dramatic increase in MISTO membership over time reflects party fragmentation and individual defections.

**Table A2. Distribution of seats in the Senato della Repubblica across the three cabinets of the 18th legislature.**
Seat counts and percentages refer to the composition of the Senato della Repubblica at the time of each cabinet’s investiture vote. Percentages are computed relative to the total number of Senate seats at each time point, excluding senators-for-life and accounting for vacancies. IV-PSI denotes the joint parliamentary group of Italia Viva and Partito Socialista Italiano; LSP-PSdAz denotes the joint group of Lega Salvini Premier and Partito Sardo d’Azione.

| Group | Conte I (June 2018) | Conte II (September 2019) | Draghi (February 2021) |
|---|---|---|---|
| M5S | 109 (34.8%) | 107 (34.2%) | 72 (23.0%) |
| LSP-PSDAZ | 58 (18.5%) | 58 (18.6%) | 62 (19.9%) |
| FI | 57 (18.2%) | 57 (18.2%) | 49 (15.6%) |
| PD | 52 (16.6%) | 51 (16.3%) | 38 (12.1%) |
| FDI | 18 (5.8%) | 18 (5.9%) | 20 (6.5%) |
| AUT | 7 (2.2%) | 7 (2.3%) | 8 (2.5%) |
| IV-PSI | – | – | 15 (4.7%) |
| MISTO | 12 (3.8%) | 14 (4.6%) | 49 (15.6%) |
| **Majority** | **166** | **173** | **268** |
| **Opposition** | **142** | **136** | **37** |

# A.5 Computational Implementation

All data processing and analysis were performed using Python 3.9 on a Linux workstation (Ubuntu 20.04, 32GB RAM, Intel i7-9700K processor). Key libraries and versions:

- **Data collection**: `requests` 2.26.0, `selenium` 3.141.0, `beautifulsoup4` 4.10.0
- **Database management**: `sqlalchemy` 1.4.23, `pymysql` 1.0.2
- **Data processing**: `pandas` 1.3.3, `numpy` 1.21.2
- **Network analysis**: `networkx` 2.6.3
- **Visualization**: `matplotlib` 3.4.3, `seaborn` 0.11.2

Network construction and community detection for a single 50-day period (approximately 160 votes, 630 Camera deputies) required approximately 45 seconds of computation time. The complete analysis across all six networks (3 cabinets × 2 chambers) executed in under 5 minutes.

The modularity optimization algorithm converged within 15-30 iterations for all networks analyzed. We verified stability by running the algorithm 10 times with different random seeds; all runs produced identical community assignments, confirming convergence to global (or at least very stable local) maxima.

# A.6 Reproducibility and Code Availability

Complete code for data collection, database construction, network analysis, and visualization will be released as an open-source repository under MIT license at [GitHub URL to be added upon publication]. The repository includes:

- SPARQL query templates for Camera and Senato endpoints
- Selenium scripts for web scraping with error handling
- SQLAlchemy model definitions matching the database schema
- Network construction functions with full documentation
- Community detection and metrics calculation implementations
- Visualization scripts reproducing all figures in the main text

README documentation provides step-by-step instructions for: 1. Setting up the database environment 2. Executing data collection (estimated time: 24-48 hours) 3. Running preprocessing and quality checks 4. Generating networks and computing metrics

5. Reproducing all analyses and figures

The code is designed for adaptability to other parliamentary systems. Key parameters (SPARQL endpoint URLs, data field names, time periods) are defined in configuration files, enabling researchers to apply the framework to different parliaments by modifying these settings rather than rewriting core logic.

## A.7 Ethical Considerations and Data Privacy

All data used in this study are already public under Italian transparency laws. Parliamentary votes are public acts, and MPs' voting records are matters of public record necessary for democratic accountability. No personal data beyond what is already published by the parliament (names, party affiliations, votes) are collected or analyzed.

Individual MP names appear in our database but are not reported in the main text except when discussing aggregate patterns (e.g., "21 MPs show high betweenness"). The focus is on structural patterns and group-level dynamics rather than individual behavior tracking.

The Italian Parliament's open data policy explicitly permits reuse of published data for research, commercial, and civic purposes. Our data collection complies with all terms of service and rate-limiting guidelines established by the parliament's technical team.